\def\la{\mathrel{\mathchoice {\vcenter{\offinterlineskip\halign{\hfil
$\displaystyle##$\hfil\cr<\cr\sim\cr}}}
{\vcenter{\offinterlineskip\halign{\hfil$\textstyle##$\hfil\cr
<\cr\sim\cr}}}
{\vcenter{\offinterlineskip\halign{\hfil$\scriptstyle##$\hfil\cr
<\cr\sim\cr}}}
{\vcenter{\offinterlineskip\halign{\hfil$\scriptscriptstyle##$\hfil\cr
<\cr\sim\cr}}}}}

\def\ga{\mathrel{\mathchoice {\vcenter{\offinterlineskip\halign{\hfil
$\displaystyle##$\hfil\cr>\cr\sim\cr}}}
{\vcenter{\offinterlineskip\halign{\hfil$\textstyle##$\hfil\cr
>\cr\sim\cr}}}
{\vcenter{\offinterlineskip\halign{\hfil$\scriptstyle##$\hfil\cr
>\cr\sim\cr}}}
{\vcenter{\offinterlineskip\halign{\hfil$\scriptscriptstyle##$\hfil\cr
>\cr\sim\cr}}}}}

\def\HCOP {\hbox{${\rm HCO}^+$}}      
\def\DCOP {\hbox{${\rm DCO}^+$}}    
\def\DTHCOP {\hbox{${\rm D^{13}CO}^+$}}    
\def\DCEIOP {\hbox{${\rm DC^{18}O}^+$}}      
\def\NTHP {\hbox{${\rm N}_2{\rm H}^+$}} 

\documentclass[referee]{aa} 
\usepackage{graphicx}
\usepackage{amsmath}
\usepackage{dcolumn}

\newcommand{\kms}{\hbox{km\,s$^{-1}$}}

\begin{document}

   \title{Laboratory and space spectroscopy of \DCOP }

   \author{Paola Caselli \inst1 \and Luca Dore\inst2 }


   \institute{INAF - Osservatorio Astrofisico
              di Arcetri, Largo E. Fermi 5, I-50125 Firenze, Italy;
              email: caselli@arcetri.astro.it \and
              Dipartimento di Chimica ``G. Ciamician'', Universit\`{a}
              di Bologna, via Selmi 2, I-40126 Bologna, Italy;
              email: luca.dore@unibo.it}


   \titlerunning{Laboratory and space spectroscopy of \DCOP }

\abstract {The rotational spectra of \DCOP and its two isotopomers \DTHCOP and \DCEIOP, produced in
a negative glow discharge cell, have been recorded in the 137--792 GHz region, which includes lines
from $J=2\leftarrow 1$ up to $J=11\leftarrow 10$. The determined rotational and centrifugal
distortion constants allow to predict the \DCOP rotational spectrum up to 1000 GHz with an accuracy
of 1 part in $10^8$ or better. This is important for kinematic studies of dense molecular cloud
cores and for future far--infrared observations. We also report on the first detection of the
hyperfine structure of the \DCOP (1--0) line made at the IRAM 30m antenna, toward the quiescent
starless cloud core L1512, in the Taurus Molecular Cloud. We point out that this is the {\it first
observation of the hyperfine splitting due to the deuteron}. This allowed us to quantify the
effects of the hyperfine splitting on the line width determination; if the hyperfine structure is
not taken into account in the line fit, the \DCOP (1--0) line width is overestimated by a
significant factor ($\ga$ 2).  } \maketitle

\section{Introduction}\

In the past few years it has been realized how important is to know with
high precision the frequencies of the molecular transitions used to
investigate the physics and the chemistry of interstellar clouds and
star forming regions (e.g. Mardones et al. 1997; Lee, Myers \& Tafalla
1999, 2001).  This is because spectral line observations are unique
tools to study internal motions of dense molecular cloud material,
within which stars will be or are forming.  In this sense, molecular
line observations are crucial to test current models of cloud core and
star formation and give insights on molecular cloud evolution.

In particular, before low-mass stars form, the progenitor (or starless) core is typically a cold
($\sim$ 10 K) and quiescent region, where the (optically thin) line widths are slightly
(approximately a factor of 2) larger than the thermal value $\Delta v_{\rm T}$ (e.g. Caselli et al.
2002).  This means that $\Delta v_{\rm obs}$ $\simeq$ 2 $\times$ $\Delta v_{\rm T}$ $\simeq$ 0.3
$(T/10 {\rm K})^{1/2}$ $(A_{\rm X}/30 \, {\rm amu})^{-1/2}$ \kms , where $T$ is the gas temperature
and $A_{\rm X}$ is the molecular weight. This already shows that to determine the LSR velocity of a
cold (T = 10 K) cloud with an accuracy comparable to the standard deviation of the corresponding
gaussian line profile ($\sigma_{\rm V_{\rm LSR}}$ = $\Delta v_{\rm obs}/\sqrt{8 ln(2)}  \la 0.1$
\kms ), for a molecular species with $A_{\rm X} \sim 30$~amu (such as CO, \HCOP , and \DCOP ), the
observed line frequency should have an uncertainty of $\Delta \nu$(kHz) $\la$ $0.1 \times 3.3
\times \nu_0$(GHz) (where $\nu_0$ is the rest frequency), or about  30 kHz at millimeter
wavelengths.

If observed lines are optically thick and if they are tracing inward motions, they appear broader
and asymmetric, typically double peaked, with the blue peak stronger than the red peak (e.g. Leung
\& Brown 1977; Zhou et al. 1990).  However, to quantify the infall velocity one needs to observe an
optically thin tracer and compare the thin and thick profiles (e.g. Myers et al. 1996).  For this
purpose, one typically uses the normalized velocity difference between the thin and thick lines
($\delta V$, as defined by Mardones et al. 1997). In this case, the condition $\delta V
/\sigma_{\delta V}$ $>$ 3 is only obtained if the error on $V$ is $\sigma_V$ $\la$0.01 \kms \
(assuming average values of $\delta V$ $\sim$ 0.2, $\sigma_{\Delta V}$ $\sim$ 0.03 \kms , and
$\Delta V$ $\simeq$ 0.3 \kms ; see Mardones et al. 1997). This implies a frequency precision of
$\sigma_{\nu}$(kHz) $\la$ 0.03$\times \nu_0$(GHz), or about 3 kHz at $\sim$100 GHz. All this
demonstrates the need of laboratory work to accurately measure line frequencies.  \cite{gmt03} have
also pointed out that it is a challenge to the spectroscopist the determination of line frequencies
with a precision large enough to identify motions in molecular cloud cores from tiny velocity
shifts, which typically  require a precision of a few parts in $10^8$. In the case of \DCOP \ this
is equivalent to 1.7\% of the Doppler line width at 77~K.

~\DCOP \ is a particularly suitable species to study the chemistry, in particular the deuterium
fractionation (e.g. Saito et al. 2002) and the electron fraction (e.g. Gu\'elin et al. 1977,
Caselli et al. 1998, Williams et al. 1998) and the kinematics of the dense cores.  The lower
rotational transitions are easily detected, as it is proved by the large amount of literature on
this molecular ion, from interstellar clouds (e.g. Wootten et al. 1982) to dense cloud cores (e.g.
Butner et al. 1995), to circumstellar disks (van Dishoeck et al. 2003).  \DCOP \ has been recently
surveyed in a large sample of starless cores by Lee et al. (2004), with the aim of detecting infall
motions. However, the previous measurements of \DCOP millimeter-wave frequencies, carried out in
two different laboratories (Bogey et al. \cite{Bogey81,Sastry81}), gave results differing up to 45
part in $10^8$ for the $J=3 \leftarrow 2$ line (see Table~\ref{comparison}); therefore there is a
need for more accurate frequency measurements of \DCOP \ lines in order to use them in the study of
star-forming molecular cloud cores. In this paper, we repeat with far better precision the old
measurements of millimeter-wave transitions of \DCOP, \DTHCOP, and \DCEIOP (Bogey at al.
\cite{Bogey81}), and we add new measurements of submillimeter-wave transitions: the spectroscopic
constants derived from the analysis of the frequency data enable to predict spectra up to 1000 GHz
with an accuracy of at least 1 part in $10^8$ for \DCOP, 5 parts in $10^8$ for \DTHCOP, and of 1
part in $10^7$ for \DCEIOP; all the isotopomers have been observed in natural abundance.

\begin{table}[htbp]
 \caption{Observed rotational transition of \DCOP (MHz)}
 \label{comparison}
 \begin{tabular}{ccD{.}{.}{3}D{.}{.}{3}D{.}{.}{0}D{.}{.}{4}D{.}{.}{3}}
 \hline \noalign{\smallskip} $J'$ & $J$ & \multicolumn{1}{c}{Bogey at
 al. \cite{Bogey81}} & \multicolumn{1}{c}{\cite{Sastry81}} &
 \multicolumn{1}{c}{diff.$^{\mathrm{a}}$/ppb} &
 \multicolumn{1}{c}{this work} &
 \multicolumn{1}{c}{radiastronomy$^{\mathrm{b}}$} \cr
\noalign{\smallskip} \hline \noalign{\smallskip}
  2 & 1 & 144\,077.342 & 144\,077.319 & 160 & 144\,077.2890 & 144\,077.321\cr
  3 & 2 & 216\,112.701 & 216\,112.604 & 449 & 216\,112.5822\cr
  4 & 3 & 288\,144.018 & 288\,143.911 & 371 & 288\,143.8583 & 288\,143.911\cr
  5 & 4 & 360\,169.980 & 360\,169.881 & 275 & 360\,169.7783 & 360\,169.830\cr
  6 & 5 &              & 432\,189.033 &     &  432\,189.0052 \cr
\noalign{\smallskip} \hline
\end{tabular}
\begin{list}{}{}
\item[$^{\mathrm{a}}$] difference between measurements of Bogey at al. \cite{Bogey81} and
of \cite{Sastry81}
\item[$^{\mathrm{b}}$]~\cite{Lovas04}
\end{list}
\end{table}

\section{Observations}

\subsection{Laboratory measurements}\

The laboratory spectrum was observed with a frequency-modulated
millimeter-wave spectrometer (\cite{C&D90a}) equipped with a
double-pass negative glow discharge cell (\cite{Dore99}) made of a
Pyrex tube 3.25 m long and 5 cm in diameter. The radiation source was
a frequency multiplier driven by a Gunn diode oscillator working in
the region 68--76 GHz (Farran Technology Limited) for the lines up to
$J=4\leftarrow 3$; for the high frequency transitions, a doubler in
cascade with a multiplier (RPG - Radiometer Physics GmbH) was driven
by Gunn oscillators working in the region 75--115 GHz (J.~E.~Carlstrom
Co and RPG) to cover frequencies up to 792 GHz. Two phase-lock loops
allow the stabilization of the Gunn oscillator with respect to a
frequency synthesizer, which is driven by a 5-MHz rubidium frequency
standard. The frequency modulation of the radiation is obtained by
sine-wave modulating with low distortion (total harmonic distortion
less than $0.01\%$) the reference signal of the wide-band
Gunn-synchronizer. The signal, detected by a liquid-helium-cooled InSb
hot electron bolometer (QMC Instr. Ltd. type QFI/2), is demodulated at
2-{\it f} by a lock-in amplifier.

$\DCOP$ was produced by flowing a 1:1 mixture of CO$_2$ and D$_2$ (1
mTorr) in Ar buffer gas with a total pressure of about 5 mTorr and
discharging with a current of a few mA; \DTHCOP and \DCEIOP were
observed in natural abundance. The cell was cooled at 77~K by liquid
nitrogen circulation, and an axial magnetic field up to about 400~G
was applied throughout the length of the discharge. With this
longitudinal magnetic field applied, ions are produced and observed in
the negative glow (\cite{DeL82}), which is a nearly field free region
and where they are expected to show negligible Doppler shift due to
the drift velocity, occurring, instead, in the positive column
(\cite{Sastry81}) where a low axial electric field is present. In
addition, the double-pass arrangement would compensate for such a
shift, if present.

A typical spectrum is recorded by sweeping the frequency up and down
(several times if signal averaging is needed) in steps of 5 kHz at a
rate of 0.8 MHz s$^{-1}$, with a lock-in amplifier time constant of 3
ms and a frequency modulation depth comparable to the Doppler width or
larger for the rarer isotopomers. Since we have full flexibility in
controlling scanning rate, number of data points and modulation depth,
the values of these parameters have been adjusted to prevent any bias
of the measured transition frequency, which is recovered from a line
shape analysis of the spectral profile (\cite{C&D90a,Dore03}).

\subsection{~\DCOP (1--0) toward the quiescent cloud core L1512}\
\label{s_l1512}

\begin{figure}[ht!]
 \centering
\resizebox{10cm}{!}{\includegraphics{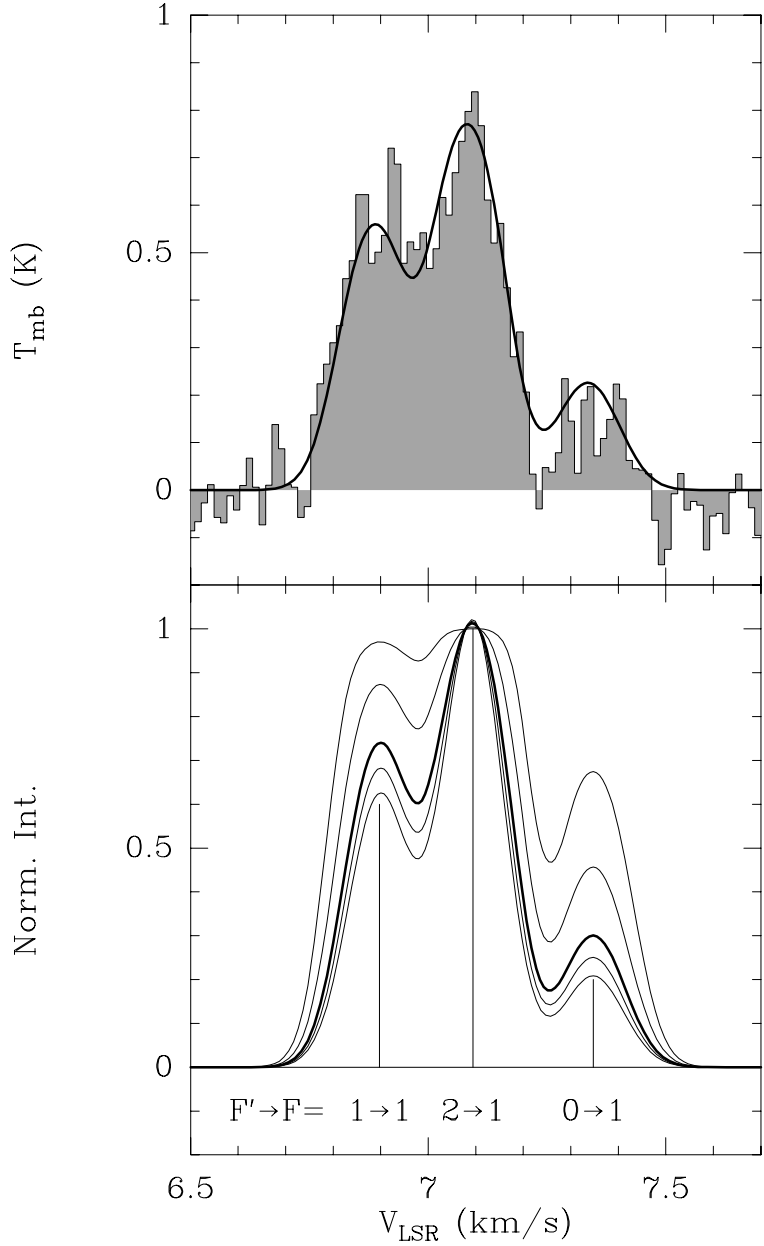}}
\caption{({\it Top panel}) Observed spectrum of \DCOP (1--0) toward the quiescent starless cloud
core L1512, in the Taurus Molecular Cloud. The units are in main beam brightness temperature,
assuming a unity filling factor. This is the first spectrum showing the hyperfine structure of the
line due to the deuteron electric quadrupole interaction.  The line is significantly broadened by
the hyperfine structure, and the width is a factor of 2.3 larger if the splitting is not taken into
account in the fit.  The black curve is the hfs fit performed in CLASS. ({\it Bottom panel})
Simulated spectrum of the \DCOP (1--0) line assuming the same parameters of the observed line
($T_{\rm ex}$ = 4 K and intrinsic $\Delta v$ = 0.15 \kms ) but with different values of $\tau_{\rm
TOT}$ (from least to most intense curves: 0.1, 1, 2 (as observed), 5, and 10. Vertical bars
indicate the positions of the three hyperfines with their relative intensities in the optically
thin limit.} \label{l1512}
\end{figure}

The \DCOP (1--0) spectrum toward the quiescent Taurus starless core L1512 (Fig.~\ref{l1512}) has
been obtained with the IRAM--30m antenna, located at Pico Veleta (Spain) in August 2004. The
adopted coordinates were R.A.(1950) = 05$^{\rm h}$ 00$^{\rm m}$ 54.4$^{\rm s}$, Dec.(1950) =
32$^{\circ}$ 39$^{\prime}$ 00.0$^{\prime\prime}$. We used the recent extension of the 3mm tuning
range below 80 GHz and carried out the observations using the frequency switching technique with a
throw of 3.9 MHz. The spectral resolution is 3.3 kHz, which corresponds to 14 m/s at 72 GHz.  Given
that below $\sim$ 75 GHz the image gain depends strongly on the tuning parameters and often varies
across the bandpass of the receiver, we measured the sideband gain ratio and corrected the observed
spectrum accordingly. The pointing was checked every 2 hours and found to be accurate within about
10$^{\prime\prime}$ because of anomalous refraction problems during very good atmospheric
conditions.  The half power beam width (HPBW) was 33$^{\prime\prime}$. The units are main beam
brightness temperature, assuming a source filling factor of unity. The beam and forward
efficiencies were 0.79 and 0.95, respectively.  The rms noise of the spectrum is 0.09 K.

Figure~\ref{l1512} clearly shows that the \DCOP (1--0) line profile is complex and contains
structure.  Indeed, the three detected features are consistent with the hyperfine splitting due to
the interaction between the molecular electric field gradient and the electric quadrupole moments
of the deuteron (spin = 1), which produces the line splitting into three components. The velocity
shifts predicted by theory are larger than the observed separations between the three components by
about 0.1 \kms , suggesting that the quadrupole constant needs to be refined.  Thus, we estimated
the separations performing a gaussian fit to the three components of the \DCOP (1--0) spectrum.
This allowed us to derive the velocity separations of -0.198$\pm$0.012 \kms \ and 0.256$\pm$0.017
\kms \ between the main (F$^{\prime}$,F = 2,1) component and the low (F$^{\prime}$,F = 1,1) and
high (F$^{\prime}$,F = 0,1) velocity components, respectively.  With this new values of the
frequency shifts and the relative intensities of the three components (3/9,5/9,1/9 for the (1,1),
(2,1) and (0,1) hyperfines, respectively) we then performed an hfs fit using the IRAM reduction
package CLASS, shown in Figure 1 (upper panel) by the thick curve. Forcing the resulting LSR
velocity ($V_{\rm LSR}$) to coincide with that measured with the high sensitivity \NTHP (1--0)
spectrum reported in Caselli et al. (1995), which has been corrected to account for the new \NTHP
(1--0) frequency value (93176.2608 MHz $\pm$ 6 kHz for the component F$_1$,F = 0,1$\rightarrow$1,2)
estimated by Dore et al. (2004), we obtain the frequency values reported in Table 2. These new
frequencies have finally been used to refine the quadrupolar constant (see next section).

The hfs fit in CLASS gives the following parameters: $V_{\rm LSR}$ = 7.094$\pm$0.004 \kms ,
intrinsic (i.e. corrected for optical depth effects) line width $\Delta v$ = 0.149$\pm$0.008 \kms ,
total optical depth (i.e. the sum of the optical depths of the three hyperfines) $\tau_{\rm TOT}$ =
1.9$\pm$0.5, excitation temperature $T_{\rm ex}$ = 4$\pm$1 K.  We note here that the observed line
width is only 1.25 times larger than the \DCOP \ thermal width $\Delta v_{\rm T}$, assuming a
kinetic temperature of 10 K (or $\Delta v$ = $\Delta v_{\rm T}$ if $T$ = 15 K).   If the hyperfine
structure is not taken into account, the \DCOP (1--0) line width obtained with a simple Gaussian
fit is a factor of 2.3 times larger, in net contrast with the narrow ($\sim$ 0.18 \kms ) \NTHP
(1--0) line width observed in the same position (Caselli et al. 1995).  Therefore, we conclude that
the differences in the widths of \DCOP (1--0) and NH$_3$(1,1) lines  observed in previous work
(e.g. compare Butner et al. 1995 with Benson \& Myers 1989) are likely due to neglecting the
hyperfine splitting.  The bottom panel of Fig.~\ref{l1512} shows how the line profile changes in
case of large optical depths. For $\tau_{\rm TOT}$ = 10, the weakest (F$^{\prime}$, F = 0,1)
component becomes bright enough to further enlarge the line width by other $\sim$0.25 \kms \ (so
that the total linewidth will be about 0.6 \kms , if the splitting is not considered in the fit).

\begin{figure}[ht!]
 \centering
\resizebox{10cm}{!}{\includegraphics{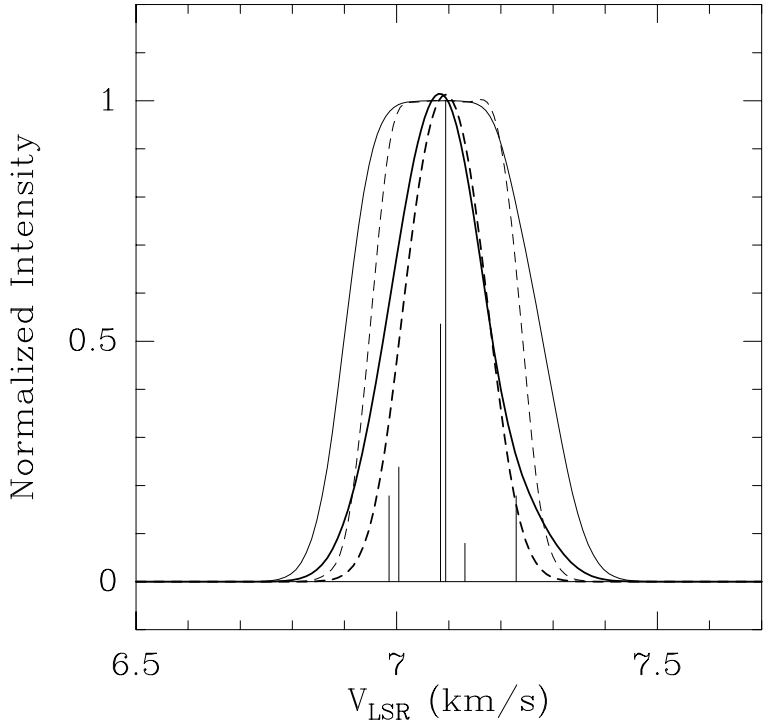}}
\caption{Simulated spectrum of the J = 2$\rightarrow$1 transition of \DCOP \ in L1512, including
the hyperfine splitting due to the deuteron electric quadrupole interaction (black curves).  The
assumed excitation temperature is 4 K and the intrinsic linewidth is 0.15 km s$^{-1}$, as found for
the \DCOP (1--0) line, whereas the total optical depth is 1 (thick curve) and 10 (thin curve).
Dashed curves are the expected \DCOP (2--1) lines in case of no hyperfine structure, again assuming
$\tau_{\rm TOT}$ = 1 (thick dashed curve) and 10 (thin dashed curve). Vertical lines mark the
velocities of the 6 hyperfine components with relative intensities given by their statistical
weight. The hyperfine splitting leads to a $\sim$20 and $\sim$30\% increase of the line width for
$\tau_{\rm TOT}$ = 1 and 10, respectively.} \label{dcop21}
\end{figure}

The effects of the hyperfine structure on the line width and profile become less important for
higher $J$ transitions.  In Fig.~\ref{dcop21}, the simulated spectrum of the \DCOP (2--1) line,
with the same parameters as the \DCOP (1--0) line in Fig.~\ref{l1512}, is shown. The six hyperfine
components are blended together, but there is still some ($\sim$20 \%) extra--broadening and a
small (0.01 km s$^{-1}$) line center shift resulting from the hfs splitting, which should be taken
into account when analysing \DCOP (2--1) profiles.  In particular, this may be important to
quantify the infall velocity in the central regions of starless cores, as recently done by Lee et
al. (2004).  On the other hand, the components of the $J$ = 3$\rightarrow$2 transition (as well as
those of higher J transitions) are heavily blended together, and the line width measurement is not
affected at all by their presence.

\section{Analysis and discussion}\
\label{s_analysis}

The measured transition frequencies of \DCOP, \DTHCOP, and \DCEIOP are
listed in Tables~\ref{parent}, \ref{C-13}, and \ref{O-18}; they are
mean values obtained from 5 to 14 measurements. The standard errors of
the mean result unrealistically small (for \DCOP they range from 0.1
to 1 kHz), therefore the uncertainties reported in the last column of
each table were estimated from the ratio of linewidth to
signal-to-noise ratio, and were roughly scaled according to the
standard errors.

The experimental data were fitted, in a weighted least-squares procedure, to the standard
expression for the frequency of the rotational transition $J+1 \leftarrow J$:
\begin{equation}
\nu_0 = 2B(J+1)-4D_J(J+1)^3+H_J(J+1)^3[(J+1)^3-J^3],
\end{equation}
where $B$ is the rotational constant and $D_J$ and $H_J$ are the quartic and sextic centrifugal
distortion constants, respectively; the weights were the inverse-square of the uncertainties. The
standard deviation $\sigma_{fit}$ of the fit for each isotopomer is just a few kHz with the
inclusion of $H_J$ as fit parameter. An assessment of the significance of the latter can be carried
out by comparison with DCN centrifugal distortion constants: the values (\cite{DCN04}) reported in
Tables~\ref{parent} and \ref{C-13} result comparable with the values of $D_J$ and $H_J$ derived in
this work for \DCOP and \DTHCOP, therefore these constants may be assumed correctly determined.

 In the case of \DCOP, the three hyperfine frequencies of the $J= 1\rightarrow 0$ transition,
accurately determined as reported above, could be used to derive the quadrupole coupling ($eqQ$)
and spin rotation ($C_I$) constants of the D nucleus (spin quantum number $I=1$) according to the
expression:
\begin{equation}
\nu_{hf} = \nu_0-eqQ [Y(J+1,I,F')-Y(J,I,F)]+C_I [C(J+1,I,F')-C(J,I,F)],
\end{equation}
where $F$ is the total angular momentum ($\vec{F}=\vec{J}+\vec{I}$) quantum number,
$C(J,I,F)=F(F+1)-I(I+1)-J(J+1)$ and the Casimir function $Y(J,I,F)$ is given by:
\begin{equation}
Y(J,I,F)]=\frac{\frac{3}{4}C(C+1)-I(I+1)J(J+1)}{2(2I-1)(2J+3)I(2I-1)}.
\end{equation}
The Pickett's SPFIT fitting program (Pickett \cite{Pick91}) allowed us to carry out simultaneously
the hfs and centrifugal analyses in a global fit of laboratory and  radioastronomical transition
frequencies, whose weights were the inverse-square of the uncertainties. For the hyperfines, their
assumed uncertainties account for both the standard errors of the hfs fit and the 6 kHz uncertainty
of the \NTHP (1--0) reference line (\cite{Dore04}).

Tables~\ref{parent}, \ref{C-13}, and  \ref{O-18} list, for each isotopomer, the experimental
frequencies with their uncertainties and the residuals of the least-squares fit carried out to
determine the spectroscopic constants reported at the bottom of the table. These accurate constants
(see next paragraphs for a discussion of accuracy) allow us to recommend the rest frequencies
reported in Table~\ref{restfr}: the hyperfine components of the $J=1 \leftarrow 0$ and $J=2
\leftarrow 1$ transitions are listed for \DCOP, while unsplit transition frequencies from $J=2$ up
to $J=14$ are reported for all three isotopomers. It is worth mentioning that the quoted
uncertainties in Table~\ref{restfr} are  $1\sigma$ errors derived from the least-squares
variance-covariance matrix for the fitted spectroscopic parameters.

\begin{table}[htbp]
 \caption{Rotational transition frequencies and spectroscopic
constants of DCO$^+$}\label{parent}
\begin{tabular}{llD{.}{.}{9}D{.}{.}{6}D{.}{.}{8}} \hline
\noalign{\smallskip} $J'(F')$ & $J(F)$ & \multicolumn{1}{c}{observed/MHz} &
\multicolumn{1}{c}{obs.-calc./kHz}
  & \multicolumn{1}{c}{uncertainties/kHz$^{\mathrm{a}}$} \cr
\noalign{\smallskip} \hline \noalign{\smallskip}
  1(0) & 0(1) &  72\,039.2413 & -0.1 & 7.3 \cr
  1(2) & 0(1) &  72\,039.3028 & -0.3 & 6.1 \cr
  1(1) & 0(1) &  72\,039.3504 & -0.2 & 6.7 \cr
  2 & 1 & 144\,077.2890^{\mathrm{b}} & 3.9 & 5 \cr
  3 & 2 & 216\,112.5822 &  2.1 & 5 \cr
  4 & 3 & 288\,143.8583 &  0.5 & 5 \cr
  5 & 4 & 360\,169.7783 & -1.2 & 5 \cr
  6 & 5 & 432\,189.0052 & -1.3 & 5 \cr
  7 & 6 & 504\,200.1999 & -0.5 & 5 \cr
  8 & 7 & 576\,202.0239 &  1.1 & 5 \cr
  9 & 8 & 648\,193.1357 &  0.0 & 5 \cr
 10 & 9 & 720\,172.2024 &  1.1 & 10 \cr
 11 &10 & 792\,137.8811 & -1.2 & 10 \cr
  \multicolumn{5}{l}{$\sigma_{fit}^{\mathrm{c}}$/kHz 1.3} \cr
  \cline{1-5} \multicolumn{2}{l}{Constant$^{\mathrm{d}}$} &
  \multicolumn{1}{c}{this work} &
  \multicolumn{1}{c}{previous$^{\mathrm{e}}$} &
  \multicolumn{1}{c}{DC$^{15}$N} \cr \cline{1-5}
  \multicolumn{2}{l}{$B$/MHz rotational} & 36\,019.76765(14) & 36\,019.784(2)\cr
  \multicolumn{2}{l}{$D_J$/kHz quartic c. d.$^{\mathrm{f}}$}
   & 55.7960(22) & 55.71(12) & 54.3924(15)^{\mathrm{g}}\cr
  \multicolumn{2}{l}{$H_J$/Hz sextic c. d.$^{\mathrm{f}}$} & 0.054(11) & & 0.0633(12)^{\mathrm{g}} \cr
  \multicolumn{2}{l}{$eqQ$/kHz quadrupole}  & 147.8(35) & & 200.6(5)^{\mathrm{h}} \cr
  \multicolumn{2}{l}{$C_I$/kHz spin rotation} & -1.59(78) & & -1.9(1)^{\mathrm{h}} \cr
\noalign{\smallskip} \hline
\end{tabular}
\begin{list}{}{}
\item[$^{\mathrm{a}}$] Uncertainties estimated as explained in the text
\item[$^{\mathrm{b}}$] Not included in the fit, see text
\item[$^{\mathrm{c}}$] ($\Sigma$(obs.-calc.)$^2$/degrees of freedom)$^{1/2}$
\item[$^{\mathrm{d}}$] Standard errors are reported in parentheses in units of the last quoted digits
\item[$^{\mathrm{e}}$] Bogey et al. \cite{Bogey81}
\item[$^{\mathrm{f}}$] centrifugal distortion
\item[$^{\mathrm{g}}$] ~\cite{DCN04}
\item[$^{\mathrm{h}}$] ~\cite{C&D90b}
\end{list}
\end{table}

\begin{table}[htbp]
 \caption{Rotational transition frequencies and spectroscopic
constants of D$^{13}$CO$^+$}\label{C-13}
\begin{tabular}{ccD{.}{.}{9}D{.}{.}{6}D{.}{.}{8}} \hline
\noalign{\smallskip} $J'$ & $J$ & \multicolumn{1}{c}{observed/MHz} &
\multicolumn{1}{c}{obs.-calc./kHz}
  & \multicolumn{1}{c}{uncertainties/kHz$^{\mathrm{a}}$} \cr
\noalign{\smallskip} \hline \noalign{\smallskip}
  2 & 1 & 141\,465.1331^{\mathrm{b}} & 3.4 & 5 \cr
  3 & 2 & 212\,194.4920 &  1.9 & 5 \cr
  4 & 3 & 282\,920.0055 &  0.3 & 5 \cr
  5 & 4 & 353\,640.3923 & -1.3 & 5 \cr
  6 & 5 & 424\,354.3718 & -1.8 & 5 \cr
  7 & 6 & 495\,060.6646 &  0.4 & 5 \cr
  8 & 7 & 565\,757.9861 &  2.0 & 5 \cr
  9 & 8 & 636\,445.0518 & -0.8 & 5 \cr
 10 & 9 & 707\,120.5885 & -0.7 & 15 \cr
 11 &10 & 777\,783.3159 &  2.3 & 50 \cr
  \multicolumn{5}{l}{$\sigma_{fit}^{\mathrm{c}}$/kHz 1.8} \cr
  \cline{1-5} \multicolumn{2}{l}{Constant$^{\mathrm{d}}$} &
  \multicolumn{1}{c}{this work} &
  \multicolumn{1}{c}{previous$^{\mathrm{e}}$}&
  \multicolumn{1}{c}{D$^{13}$C$^{15}$N$^{\mathrm{f}}$}\cr \cline{1-5}
  \multicolumn{2}{l}{$B$/MHz rotational} & 35\,366.70968(21) &
  35\,366.712(11)\cr \multicolumn{2}{l}{$D_J$/kHz quartic
  c. d.$^{\mathrm{g}}$} & 53.4083(40) & 53.20(53) & 52.26535(96)\cr
  \multicolumn{2}{l}{$H_J$/Hz sextic c. d.$^{\mathrm{g}}$} & 0.048(24)
  & & 0.06074(73)\cr
\noalign{\smallskip} \hline
\end{tabular}
\begin{list}{}{}
\item[$^{\mathrm{a}}$] Uncertainties estimated as explained in the text
\item[$^{\mathrm{b}}$] Not included in the fit, see text
\item[$^{\mathrm{c}}$] ($\Sigma$(obs.-calc.)$^2$/degrees of freedom)$^{1/2}$
\item[$^{\mathrm{d}}$] Standard errors are reported in parentheses in units of the last quoted digits
\item[$^{\mathrm{e}}$] Bogey et al. \cite{Bogey81}
\item[$^{\mathrm{f}}$]~\cite{DCN04}
\item[$^{\mathrm{g}}$] centrifugal distortion
\end{list}
\end{table}

\begin{table}[htbp]
 \caption{Rotational transition frequencies and spectroscopic
constants of DC$^{18}$O$^+$}\label{O-18}
\begin{tabular}{ccD{.}{.}{9}D{.}{.}{6}D{.}{.}{5}} \hline
\noalign{\smallskip} $J'$ & $J$ & \multicolumn{1}{c}{observed/MHz} &
\multicolumn{1}{c}{obs.-calc./kHz}
  & \multicolumn{1}{c}{uncertainties/kHz$^{\mathrm{a}}$} \cr
\noalign{\smallskip} \hline \noalign{\smallskip}
  2 & 1 & 137\,653.5239^{\mathrm{b}} & 3.1 & 10 \cr
  3 & 2 & 206\,477.2402 & -0.9 & 5 \cr
  4 & 3 & 275\,297.3150 &  1.5 & 5 \cr
  5 & 4 & 344\,112.5213 & -1.1 & 5 \cr
  6 & 5 & 412\,921.6530 &  0.4 & 5 \cr
  7 & 6 & 481\,723.4893 &  0.3 & 5 \cr
  8 & 7 & 550\,516.8156 & -1.7 & 10 \cr
  9 & 8 & 619\,300.4281 &  4.4 & 35 \cr
 10 & 9 & 688\,073.0967 &  1.9 & 45 \cr
  \multicolumn{5}{l}{$\sigma_{fit}^{\mathrm{c}}$/kHz 2.5}
  \cr \cline{1-5} \multicolumn{2}{l}{Constant$^{\mathrm{d}}$} &
  \multicolumn{1}{c}{this work} &
  \multicolumn{1}{c}{previous$^{\mathrm{e}}$}\cr \cline{1-5}
  \multicolumn{2}{l}{$B$/MHz rotational} & 34\,413.78556(18) &
  34\,413.798(1)\cr \multicolumn{2}{l}{$D_J$/kHz quartic
  c. d.$^{\mathrm{f}}$} & 50.6704(44) & 50.29(4)\cr
  \multicolumn{2}{l}{$H_J$/Hz sextic c. d.$^{\mathrm{f}}$} & 0.109(34)
  & \cr
\noalign{\smallskip} \hline
\end{tabular}
\begin{list}{}{}
\item[$^{\mathrm{a}}$] Uncertainties estimated as explained in the text
\item[$^{\mathrm{b}}$] Not included in the fit, see text
\item[$^{\mathrm{c}}$] ($\Sigma$(obs.-calc.)$^2$/degrees of freedom)$^{1/2}$
\item[$^{\mathrm{d}}$] Standard errors are reported in parentheses in units of the last quoted digits
\item[$^{\mathrm{e}}$] Bogey et al. \cite{Bogey81}
\item[$^{\mathrm{f}}$] centrifugal distortion
\end{list}
\end{table}

As for the accuracy of the measured transition frequencies, the
sources of systematic error are of two types: those due to the
experimental set-up and procedures, and those related to the physics
of molecules in the experimental environment.

\begin{figure}[ht!]
 \centering
\includegraphics{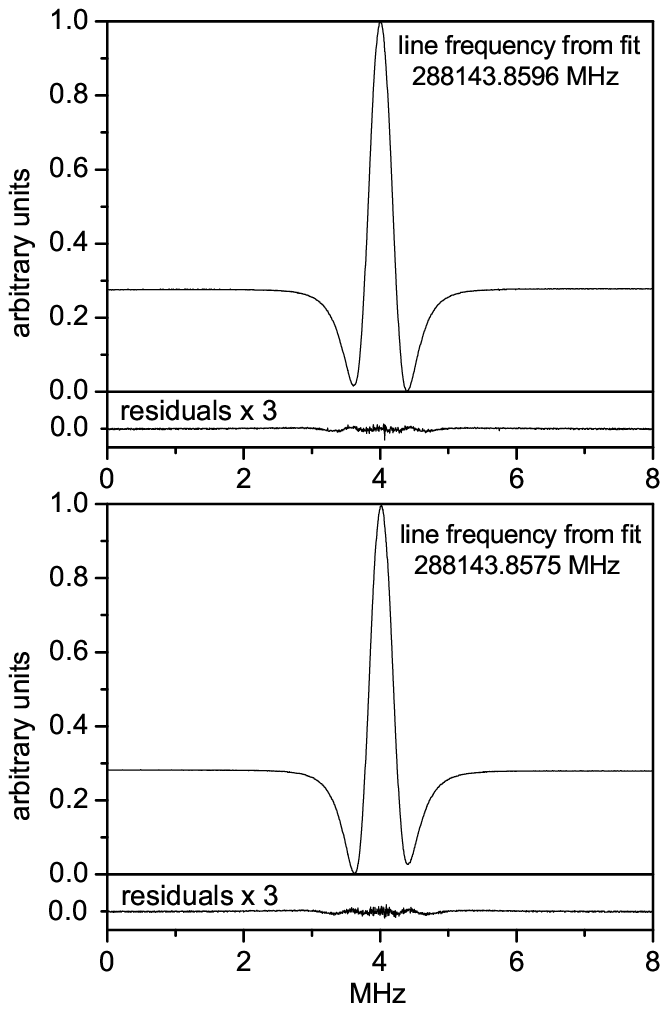}
\caption{Spectra of the $J=4 \leftarrow 3$ transition of DCO$^+$ with opposite line asymmetry due
to different standing wave pattern in the cell: the 1601 points of each spectrum were recorded for
a total time of 39 s at a rate of 0.8 MHz s$^{-1}$ with a lock-in time constant of 1 ms. Their
profile has been fit to a model Galatry profile with the inclusion of a dispersion term to recover
the line frequency.} \label{Asym}
\end{figure}

The sources of the first kind include: accuracy of the frequency standard, harmonic distortion of
the frequency modulation, integration due to slow lock-in time constant, asymmetry in the line
shape due to etalon effect in the absorption cell. The well tested performances of the spectrometer
(\cite{C&D90a,PCD02}) and a judicious choice of the lock-in time constant (less than 1/100 of the
rising time during the line recording) allowed all these error sources, except the last, to be
rendered negligible: Figure~\ref{Asym} shows, in fact, a pair of spectra of the $J= 5\leftarrow 4$
transition of \DCOP with opposite asymmetry. The shift of the apparent center frequency, however,
may be accounted for by a full profile analysis with a polynomial function describing the baseline
and a dispersion term included; this procedure gives an accurate line center (\cite{Dore03}).

The second kind of error in the line center includes: Doppler shift due to ion drift in the
discharge, gas flow through the cell due to pumping, pressure shift. The first two causes of line
shift should be negligible in the present experimental conditions, and are, anyway, largely
suppressed by the double-pass arrangement. As for the frequency shift by Ar pressure (about 4 mTorr
in this experiment), Buffa et al. (\cite{Buffa94}) have shown in the case of \HCOP that it is most
significant for the $J=1 \leftarrow 0$ transition; an estimate for \DCOP broadened by Ar
(\cite{Buffa04}) indicates that, among the rotational lines considered here, only the $J=2
\leftarrow 1$ center frequency should be slightly shifted (about 3 kHz); therefore it has been
excluded from the fit.

In the present case, there are also the hyperfine contributions of the D and $^{13}$C nuclei which
may affect the accuracy of the determined transition frequencies. These effects are more
significant for the lower $J$ transitions, therefore the most affected among those included in the
fit should be the $J=3 \leftarrow 2$. Its hf structure was predicted assuming the hf constants from
the present work for D and from Schmid-Burgk et al. (\cite{Schmid04}) for $^{13}$C; then, to
simulate the blended experimental profile, modulated Voigt functions (see~\cite{Dore03}) were
summed at each hyperfine frequency with the relative intensity as weighting factor; finally, the
center frequency of the synthetic profile was measured to check for a shift with respect to the
assumed unperturbed $J=3 \leftarrow 2$ transition frequency. From this procedure carried out for
all three isotopomers, it turned out that the apparent line center is shifted at most by 0.3 kHz
from the unperturbed center frequency: this indicates that the hyperfine structure is irrelevant in
this case, and {\it a fortiori} for the higher $J$ transitions, as seen in Sect.~\ref{s_l1512}.

\begin{table}[htbp]
 \caption{Recommended line frequencies$^{\mathrm{a}}$ (MHz) of DCO$^+$, D$^{13}$CO$^+$, and DC$^{18}$O$^+$}\label{restfr}
\begin{tabular}{llD{.}{.}{8}D{.}{.}{8}D{.}{.}{8}} \hline
\noalign{\smallskip} $J'(F') $ & $J(F)$ & \multicolumn{1}{c}{DCO$^+$} &
 \multicolumn{1}{c}{D$^{13}$CO$^+$} & \multicolumn{1}{c}{DC$^{18}$O$^+$} \cr
\noalign{\smallskip} \hline \noalign{\smallskip}
  1(0) &  0(1) &   72\,039.2414(18)    \cr
  1(2) &  0(1) &   72\,039.3031(9)    \cr
  1(1) &  0(1) &   72\,039.3506(14)    \cr
  2(1) &  1(1) &  144\,077.2144(18)    \cr
  2(1) &  1(2) &  144\,077.2619(29)    \cr
  2(3) &  1(2) &  144\,077.2804(10)    \cr
  2(2) &  1(1) &  144\,077.2851(5)    \cr
  2(1) &  1(0) &  144\,077.3237(15)    \cr
  2(2) &  1(2) &  144\,077.3326(23)    \cr
  3 &  2 &  216\,112.5800(6)   &  212\,194.4901(9)   &  206\,477.2411(7)   \cr
  4 &  3 &  288\,143.8577(6)   &  282\,920.0052(9)   &  275\,297.3135(6)   \cr
  5 &  4 &  360\,169.7795(6)   &  353\,640.3936(8)   &  344\,112.5224(6)   \cr
  6 &  5 &  432\,189.0066(5)   &  424\,354.3736(7)   &  412\,921.6526(6)   \cr
  7 &  6 &  504\,200.2004(6)   &  495\,060.6642(8)   &  481\,723.4890(8)   \cr
  8 &  7 &  576\,202.0228(7)   &  565\,757.9841(9)   &  550\,516.8173(13)   \cr
  9 &  8 &  648\,193.1356(8)   &  636\,445.0526(11)   &  619\,300.4237(31)  \cr
 10 &  9 &  720\,172.2013(11)   &  707\,120.5892(28)   &  688\,073.0947(72)  \cr
 11 & 10 &  792\,137.8823(22)   &  777\,783.3135(67)  &  756\,833.618(14)  \cr
 12 & 11 &  864\,088.8419(47)  &  848\,431.946(13)  &  825\,580.782(26)  \cr
 13 & 12 &  936\,023.7433(88)  &  919\,065.206(24)  &  894\,313.375(43)  \cr
 14 & 13 & 1\,007\,941.251(15)  &  989\,681.816(40) &  963\,030.188(68)  \cr
 15 & 14 & 1\,079\,840.028(24)  & 1\,060\,280.497(62) & 1\,031\,730.011(103) \cr
\noalign{\smallskip} \hline
\end{tabular}
\begin{list}{}{}
\item[$^{\mathrm{a}}$] The $1\sigma$ uncertainties are reported in parentheses in units of the last quoted digits
\end{list}
\end{table}

\section{Conclusions}

The present paper has shown how radioastronomical observations together with laboratory work are
needed to determine with high precision the spectroscopic parameters of molecular species.  In the
laboratory, it has been possible to measure the frequencies of \DCOP , \DTHCOP , and \DCEIOP \
lines up to 792 GHz with a fairly high accuracy. New values of the rotational, quartic, and sextic
distortion constants have been determined for the three isotopomers with such a precision to allow
to predict spectra up to 1000 GHz with an accuracy of at least 1 part in $10^8$ for \DCOP, 5 parts
in $10^8$ for \DTHCOP, and of 1 part in $10^7$ for \DCEIOP. Previous frequency estimates were much
less accurate and did not allow to make accurate kinematic studies in dense molecular cloud cores.

Using the IRAM--30m antenna, we detected for the first time the hyperfine structure of the \DCOP
(1--0) line toward the quiescent starless core L1512, in Taurus.  This is the first time that the
hyperfine structure due to a deuteron has ever been observed; this gave us the possibility of
obtaining the hyperfine parameters of \DCOP. The \DCOP (1--0) frequency has also been estimated
comparing the observed line with a high sensitivity spectrum of \NTHP (1--0), and using the new
value of the \NTHP (1--0) line as calculated by Dore et al. (2004). This value is however still
uncertain by 6 kHz because of astronomical errors, and new sub--Doppler laboratory work on \NTHP \
is planned to improve this precision. The estimated \DCOP (1--0) hf frequencies were included with
their uncertainties in a weighted least-squares fit of the laboratory data to derive accurate
spectroscopic constants. From the fitted constants, the hyperfines have been reestimated with an
accuracy of better than 2.5 parts in $10^8$.

The hyperfine structure of the \DCOP (1--0) line needs to be taken into account in the analysis of
astronomical data, to avoid overestimates of the line width by more than a factor of 2 and for a
correct interpretation of the line profile.

\begin{acknowledgements}
This work was supported by the Italian Ministry of Public Instruction, University and Research and
by ASI (contract I/R/044/02).  It is a pleasure to thank Nuria Marcelino from IRAM, who kindly
helped us during observations, and the referee, Mario Tafalla, for making useful comments which
helped to clarify some aspects of the paper. We also thank Malcolm Walmsley for his critical
reading of the manuscript.
\end{acknowledgements}

\end{document}